# ProtoGTX: Prototype-Aligned Graph Transformer for Explainable Whole-Slide Image Analysis

Iori Azuma[1], Hiroyuki Kusuhara[1], Tadahaya Mizuno[1, 2, †]

[1] Laboratory of Molecular Pharmacokinetics, Graduate School of Pharmaceutical Sciences, The University of Tokyo, Tokyo, Japan

[2] The Institute of Statistical Mathematics, Research Organization of Information and Systems, Tachikawa, Japan

tadahaya@gmail.com

† Corresponding author

**Abstract.** Deep learning has achieved strong performance in computational pathology, but explaining whole-slide image (WSI) predictions remains challenging. Existing explanation methods often highlight important regions without identifying their morphological patterns, leaving a gap between patch-level computational evidence and slide-level pathological interpretation. We propose ProtoGTX, a graph-based framework that combines spatial contextualization with a flexible morphological prototype vocabulary. ProtoGTX separates the reference prototypes used to define morphological concepts from the predictive model: the reference vocabulary can be generated from training data by unsupervised clustering or supplied externally, and remains fixed during classifier training. Patch features are first contextualized over a spatial graph, after which learnable prototypes representing morphological concepts are optimized through cross-attention. Their slide-level activation profile is aligned with the morphological composition induced by the fixed reference prototype bank. On the CPTAC and BRACS datasets, ProtoGTX outperforms the evaluated MIL, prototype-based, and graph-based baselines, with particularly pronounced improvements on BRACS. Qualitative analyses further show that spatial contextualization transforms fragmented patch-level prototype assignments into more coherent tissue regions. In BRACS, prototype maps also co-localize with pathologist annotations and regions relevant to slide-level prediction. These results demonstrate that ProtoGTX provides complementary semantic and spatial explanations while retaining a prototype vocabulary that can be defined independently of the predictive model. Code is available at https://github.com/mizuno-group/ProtoGTX.

## Introduction

Histopathology is the gold standard for cancer diagnosis. The digitization of tissue slides into Whole Slide Images (WSIs) has enabled deep learning models to achieve remarkable success in diagnostic tasks. However, the "black box" nature of these models can make their internal logic difficult to interpret in clinical research settings.

To support informed professional assessment, there is a growing need for transparent explanations that clarify the underlying rationale of AI-assisted outputs. Existing interpretability methods, such as attention scores, primarily highlight *where* the model focuses. They fail to explain *what* morphological features underlie the prediction. Bridging this gap requires models to provide insights based on biological concepts familiar to pathologists.

Pathological findings are often assessed at the whole-slide level, whereas computational models process WSIs as smaller patches for computational tractability. Multiple Instance Learning (MIL) is therefore widely used for WSI analysis, treating each slide as a bag of patches without requiring fine-grained annotations[1–6]. To provide rationales for their predictions, conventional MIL models typically rely on intermediate attention scores to identify and highlight the most salient tissue regions. However, these purely spatial explanations are fundamentally limited and lack explicit biological meaning.

To address the lack of structural context, recent studies have increasingly turned to graph-based approaches to process WSIs [7–12]. Zheng et al. [9] integrated patch-level graph representations of WSIs into a Vision Transformer architecture. Hou et al. [10] explicitly modeled multiscale spatial dependencies through resolution-aware heterogeneous graphs. Li et al. [12] introduced a dynamic approach that conceptualizes the WSI as a structured knowledge graph. While these methods successfully incorporate spatial context, a major limitation remains. The features stored in graph nodes are still high-dimensional vectors that do not directly correspond to specific morphological patterns. Even with advanced graph structures, it is difficult to identify exactly which visual concepts the model is using for its prediction. This gap makes it hard for pathologists to verify the clinical basis of the results.

Prototype-based approaches provide an alternative route toward semantically structured WSI representations by summarizing patch features using representative morphological patterns [13–16]. More recently, PTCMIL jointly learns slide-specific, task-aware prototype representations with downstream prediction through prompt-token clustering[17].

Recent studies have further integrated prototypes with task-aware aggregation or structured representations. StructMIL incorporates graph-derived tissue topology into prototype-based multiple-instance learning, with prototypes initialized by K-means and subsequently optimized for downstream prediction [18]. ProtoSG-Net instead employs non-parametric dataset-level prototypes to guide representative patch selection and semantic graph construction [19]. These studies show that combining prototypes with spatial or structural modeling is itself no longer sufficient to distinguish a new framework. A complementary question therefore remains: can the morphological vocabulary be defined independently of the predictive model, and can spatial context refine how local tissue regions are aligned to that fixed vocabulary?

To address this question, we propose ProtoGTX, a graph-based framework for spatially contextualized alignment to a morphological reference vocabulary. Our main contributions are threefold. (1) We decouple the morphological reference vocabulary from the predictive model. Reference prototypes can be obtained by unsupervised clustering of training embeddings or supplied from an external source, enabling the semantic vocabulary to be specified independently of classifier optimization. (2) We propose a Prototypical Cross-Attention module that reconstructs graph features as combinations of learnable, interpretable prototypes, whose activation profiles are aligned with the fixed reference prototype composition. (3) ProtoGTX provides complementary explanations of what morphological concepts are represented and where predictive evidence is localized, and we evaluate these properties on CPTAC and BRACS.

## Methods

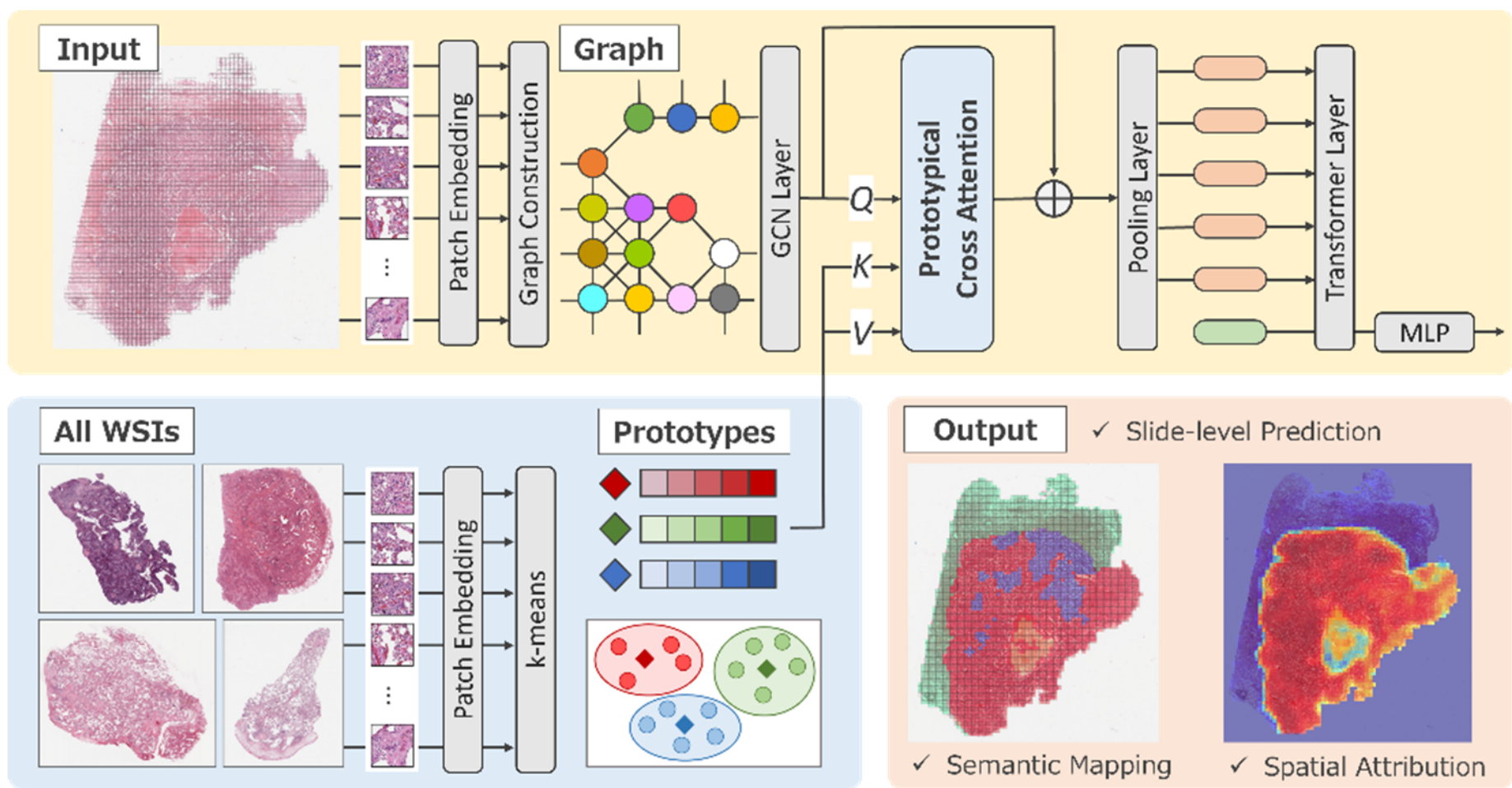

**Fig. 1.** Overview of ProtoGTX. A fixed morphological reference vocabulary is constructed from training data by K-means clustering or can be supplied externally. Reference prototypes are used to derive slide-level prototype compositions. In parallel, patch features are spatially contextualized by graph convolution and interact with learnable prototypes through cross-attention. The aggregated prototype activations are aligned with the fixed reference prototype composition, while the contextualized features are subsequently pooled and processed for slide-level classification.

### Graph Representation

Following Zheng et al. [9], we represent each WSI as an undirected graph $\mathcal{G} = (\mathcal{V}, \mathcal{E})$. The tissue regions are first partitioned into $N$ non-overlapping patches, forming the node set $\mathcal{V} = \{v_1, \dots, v_N\}$. A feature extractor encodes each patch into a $d$-dimensional embedding $\mathbf{X} \in \mathbb{R}^{N \times d}$. The edge set $\mathcal{E}$ defines spatial relationships based on 8-neighbor connectivity, represented as a binary adjacency matrix $\mathbf{A} \in \{0, 1\}^{N \times N}$, where $A_{ij} = 1$ if patches $v_i$ and $v_j$ are spatially adjacent, and $A_{ij} = 0$ otherwise.

## Reference Prototype Vocabulary

To define an interpretable morphological vocabulary independently of the predictive model, we introduce a fixed reference prototype bank $P^{ref} = \{p_1^{ref}, \dots, p_K^{ref}\}$. In our default data-driven setting, patch-level embeddings are sampled from the training WSIs and clustered using K-means, with the resulting centroids used as reference prototypes. The prototype bank is computed before classifier training and remains fixed thereafter. Importantly, ProtoGTX does not require the reference prototypes to be generated by K-means; an externally defined prototype bank can be used provided that the prototypes are represented in the same feature space.

For each patch embedding $\mathrm{x_i}$, we identify the closest reference prototype according to cosine similarity. Aggregating these assignments across a WSI yields a $K$-dimensional reference prototype composition $e$, which represents the morphological composition of the slide and serves as the semantic alignment target during ProtoGTX training.

## Prototypical Concept Alignment via Cross Attention

To map high dimensional embeddings to interpretable concepts, we first apply Graph Convolutional (GC) layers for spatial aggregation. Initializing node features as $\mathbf{H}^{(0)} = \mathbf{X}$, the $l$-th layer updates $\mathbf{H}^{(l)} \in \mathbb{R}^{N\times D}$ via:

$$\mathbf{H}^{(l+1)} = \mathrm{ReLU}\big(\widehat{\mathbf{A}}\,\mathbf{H}^{(l)}\,\mathbf{W}^{(l)}\big), \tag{1}$$

where $\mathbf{W}^{(l)}$ is a learnable filter for the $l$-th layer. The structural information is encoded by the symmetric normalized adjacency matrix $\widehat{\mathbf{A}} = \widetilde{\mathbf{D}}^{-\frac{1}{2}}\widetilde{\mathbf{A}}\widetilde{\mathbf{D}}^{-\frac{1}{2}}$, where $\widetilde{\mathbf{A}} = \mathbf{A} + \mathbf{I}$ and $\widetilde{\mathbf{D}}$ is its diagonal degree matrix. We stack multiple GC layers to capture multi-hop dependencies, yielding the spatially encoded features $\mathbf{H}_{graph} \in \mathbb{R}^{N\times D}$.

To interpret these spatially contextualized patch features, we introduce a prototypical concept alignment module. Following the cross-attention formulation of Concept Transformer[20], we define a learnable prototype $P$, where its rows represent distinct semantic concepts. The GCN-aggregated features $H$ serve as queries, while $P$ acts as both keys and values. Following the multi-head design, inputs are projected into $h$ subspaces. For each head $i \in \{1, \dots, h\}$, we compute:

$$\mathbf{Q}_i = \mathbf{H}\mathbf{W}_Q^{(i)}, \quad \mathbf{K}_i = \mathbf{P}\mathbf{W}_K^{(i)}, \quad \mathbf{V}_i = \mathbf{P}\mathbf{W}_V^{(i)}, \tag{2}$$

where $\mathbf{W}_Q^{(i)}, \mathbf{W}_K^{(i)}, \mathbf{W}_V^{(i)} \in \mathbb{R}^{D\times d_m}$ are learnable projection matrices. The alignment is performed by computing the attention weights between patches and prototypes:

$$\mathbf{S}_i = \mathrm{softmax}\left(\frac{1}{\sqrt{d_m}}\mathbf{Q}_i\mathbf{K}_i^{\top}\right) \tag{3}$$

Here, $\mathbf{S}_i \in \mathbb{R}^{N\times C}$ quantifies the affinity of each patch to specific prototypes in the $i$-th subspace. The prototype-aligned features $\mathbf{H}_{proto}$ are then synthesized by aggregating the concept values weighted by the attention scores $\mathbf{S}_i$, concatenating across all heads, and projecting back to the feature dimension:

$$\mathbf{H}_{proto} = \text{Concat}(\mathbf{S}_1\mathbf{V}_1, \ldots, \mathbf{S}_h\mathbf{V}_h)\mathbf{W}_o, \tag{4}$$

where $\mathbf{W}_o \in \mathbb{R}^{D\times D}$ is a learnable projection matrix that fuses the information from different attention heads. Finally, we employ a residual connection $\mathbf{H}_{update} = \mathbf{H}_{graph} + \mathbf{H}_{proto}$ to integrate the semantic concepts while preserving the instance-specific structural context. Additionally, the average score matrix $\bar{\mathbf{S}} = \frac{1}{h}\sum_i \mathbf{S}_i$ serves as an interpretable map, indicating the global contribution of each prototype to the WSI.

### Graph Transformer Blocks

Following Zheng et al. [9], we apply a transformer-based architecture to capture long-range dependencies among tissue clusters. Given node features $\mathbf{H}_{pool} \in \mathbb{R}^{K\times D}$ from the MinCut pooling layer, we prepend a learnable class token $\mathbf{x}_{class} \in \mathbb{R}^{1\times D}$ to construct the input sequence:

$$\mathbf{Z}_0 = [\mathbf{x}_{class}; \mathbf{H}_{pool}] \in \mathbb{R}^{(K+1)\times D}. \tag{5}$$

The $\mathbf{Z}_0$ is processed by $L$ standard transformer layers, utilizing Multi-Head Self-Attention (MSA), Multilayer Perceptron (MLP), and pre-activation Layer Normalization (LN). The update rule for the $L$ layer is:

$$\mathbf{Z}'_l = \text{MSA}(\text{LN}(\mathbf{Z}_{l-1})) + \mathbf{Z}_{l-1} \tag{6}$$

$$\mathbf{Z}_l = \text{MLP}(\text{LN}(\mathbf{Z}'_l)) + \mathbf{Z}'_l \tag{7}$$

Finally, the slide-level prediction $\hat{y}$ is derived from the output class token $\mathbf{z}_L^{(0)}$ via a $\hat{y} = \text{Classifier}\left(\text{LN}\left(\mathbf{z}_L^{(0)}\right)\right)$.

### Multi-Objective Optimization

ProtoGTX framework is optimized through a multi-objective loss function $L_{total} = L_{cls} + \lambda L_{expl}$ that balances diagnostic accuracy with semantic interpretability. The classification term $L_{cls}$ employs cross-entropy to align slide-level predictions. To bridge the semantic gap, the explanation term $L_{expl}$ minimizes the Mean Squared Error between normalized learnable-prototype activations and the reference prototype composition derived from the fixed prototype bank.

## Experiments and Results

### Datasets and Baselines

To evaluate the subtyping performance of our proposed model, we conducted comparative experiments on two widely recognized benchmark datasets: CPTAC [21] and BRACS [22]. The CPTAC dataset comprises 2,082 WSIs of lung tissue, consisting of Normal, Lung Adenocarcinoma (LUAD), and Lung Squamous Cell Carcinoma (LSCC). For BRACS dataset, we focused on the coarse-grained three-class classification task: Benign, Atypical, and Malignant. Following the protocol by Ma et al [23], we employed 5-fold cross-validation for both datasets, which partitions the data into training, validation and testing sets with a ratio of 8:1:1.

We compared our method to seven state-of-the-art methods, including CLAM-SB/MB [1], TransMIL [2], PAMIL [13], PANTHER [14], GTP [9] and WiKG [12].

### Implementation Details

We extracted features from non-overlapping patches from WSIs. To maintain a consistent physical field of view across different magnifications, we set the patch size to 512×512 pixels at 20× magnification for CPTAC, and 1024×1024 pixels for BRACS dataset (derived from multi-magnification data). Feature extraction was performed using ResNet-50 [24] pre-trained on ImageNet and UNI2 [25], a foundation model pre-trained on large-scale histopathology datasets.

### Comparison of Subtyping Performance

Following standard settings, we employed balanced accuracy and weighted F1-score as evaluation metrics. We report the mean and standard deviation to provide a reliable assessment that is robust to class imbalance.

Our proposed model consistently outperforms current state-of-the-art methods. Notably, the performance improvement is particularly pronounced on the BRACS dataset, which presents a relatively more challenging classification task, thereby demonstrating the robustness of our approach.

**Table 1.** Subtyping prediction on two datasets.

| Model | BRACS | | CPTAC | |
|---|---|---|---|---|
| | Bal. acc. | F1 | Bal. acc. | F1 |
| | | ***ResNet-50*** | | |
| CLAM-SB[1] | 0.583±0.030 | 0.660±0.033 | 0.913±0.011 | 0.913±0.011 |
| CLAM-MB[1] | 0.617±0.061 | 0.694±0.060 | 0.918±0.013 | 0.919±0.012 |
| TransMIL[2] | 0.612±0.069 | 0.689±0.069 | 0.909±0.015 | 0.911±0.014 |
| PAMIL[13] | 0.531±0.040 | 0.589±0.052 | 0.880±0.025 | 0.882±0.024 |
| PANTHER[14] | 0.502±0.043 | 0.555±0.059 | 0.908±0.012 | 0.909±0.012 |
| GTP[9] | 0.662±0.038 | 0.757±0.055 | 0.933±0.013 | 0.933±0.012 |
| WiKG[12] | 0.630±0.029 | 0.710±0.026 | 0.927±0.016 | 0.927±0.016 |
| **ProtoGTX** | **0.673±0.074** | **0.761±0.062** | **0.942±0.008** | **0.942±0.008** |
| | | ***UNI2*** | | |
| CLAM-SB[1] | 0.726±0.034 | 0.788±0.025 | 0.981±0.007 | 0.981±0.008 |
| CLAM-MB[1] | 0.716±0.065 | 0.777±0.051 | 0.984±0.006 | **0.985±0.006** |
| TransMIL[2] | 0.649±0.039 | 0.729±0.036 | 0.983±0.006 | 0.984±0.006 |
| PAMIL[13] | 0.682±0.076 | 0.746±0.050 | 0.977±0.010 | 0.977±0.010 |
| PANTHER[14] | 0.594±0.040 | 0.676±0.046 | 0.977±0.006 | 0.977±0.006 |
| GTP[9] | 0.744±0.049 | 0.816±0.021 | 0.977±0.011 | 0.977±0.011 |
| WiKG[12] | 0.715±0.057 | 0.787±0.040 | 0.977±0.016 | 0.977±0.015 |
| **ProtoGTX** | **0.816±0.062** | **0.846±0.023** | **0.985±0.009** | 0.985±0.009 |

**Spatially Guided Interpretability**

To evaluate the graph-based propagation, we visualized patch-level prototype assignments per WSI (Fig. 2A). While the baseline exhibited a fragmented, mosaic-like map intermixing noise and tissue features, our approach leverages neighbor interactions to suppress minor, non-diagnostic artifacts (e.g., C1, C2, C14, C15; Fig. 2B). Simultaneously, the model consolidates dominant prototypes into continuous, segmentation-like regions without requiring pixel-level annotations, demonstrating its capability to autonomously discover tissue architecture.

Furthermore, prototype distributions across CPTAC subtypes (Fig. 2C) revealed distinct lineage-specific patterns: C5, C9, and C13 were predominantly associated with normal tissue, C10 and C12 enriched in LUAD, and C7 and C16 frequently observed in LSCC.

Collectively, these findings confirm that integrating spatial context is essential. By filtering noise, our framework achieves WSI representations deeply grounded in pathologically meaningful morphology.

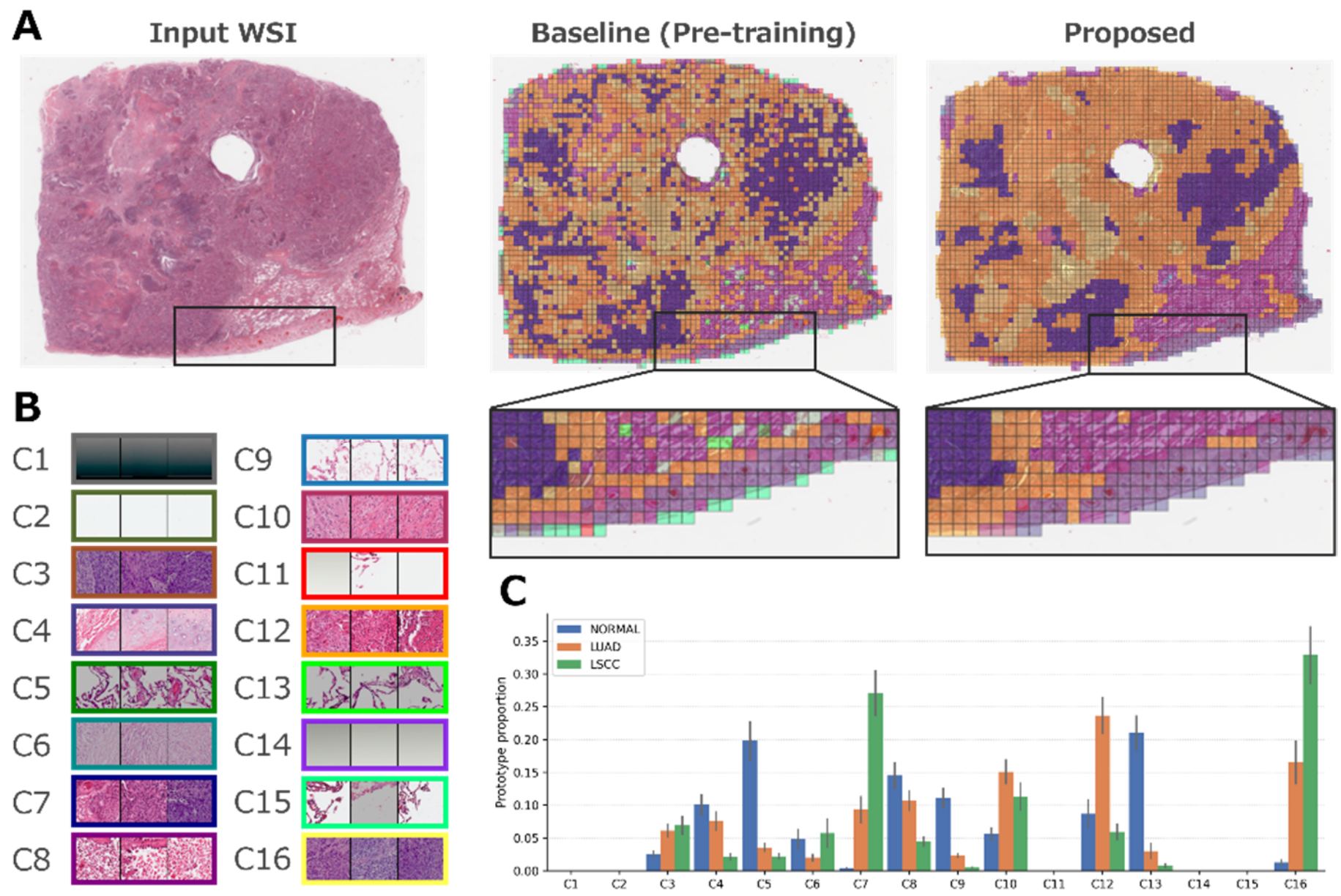

**Fig. 2.** Spatially guided prototype assignment. (A) Patch-level prototype mapping on a WSI. Compared to the fragmented baseline, the proposed model leverages spatial context to consolidate assignments into continuous, segmentation-like regions. (B) Visual dictionary of the 16 prototypes (C1–C16). (C) Distribution of prototype proportions across three classes (Normal, LUAD, and LSCC), highlighting lineage-specific morphological patterns.

## Clinically Grounded Interpretability

To examine whether the morphological vocabulary remains meaningful in an independent spatial validation setting, we used a seven-class reference prototype bank representing stages of breast cancer progression. The prototype bank was constructed independently of the WSIs used for segmentation evaluation and remained fixed during the analysis. In a representative malignant sample, the Invasive Carcinoma (IC) prototype explicitly delineates the tumor bed (Fig 3A).

Notably, this morphological mapping highly co-localizes with the regions driving the slide-level malignancy prediction. These predictive regions further overlap with the official BRACS pathologist annotations for IC [22].

The qualitative co-localization among assigned prototypes, predictive attribution, and pathologist annotation supports the spatio-semantic interpretability of ProtoGTX.

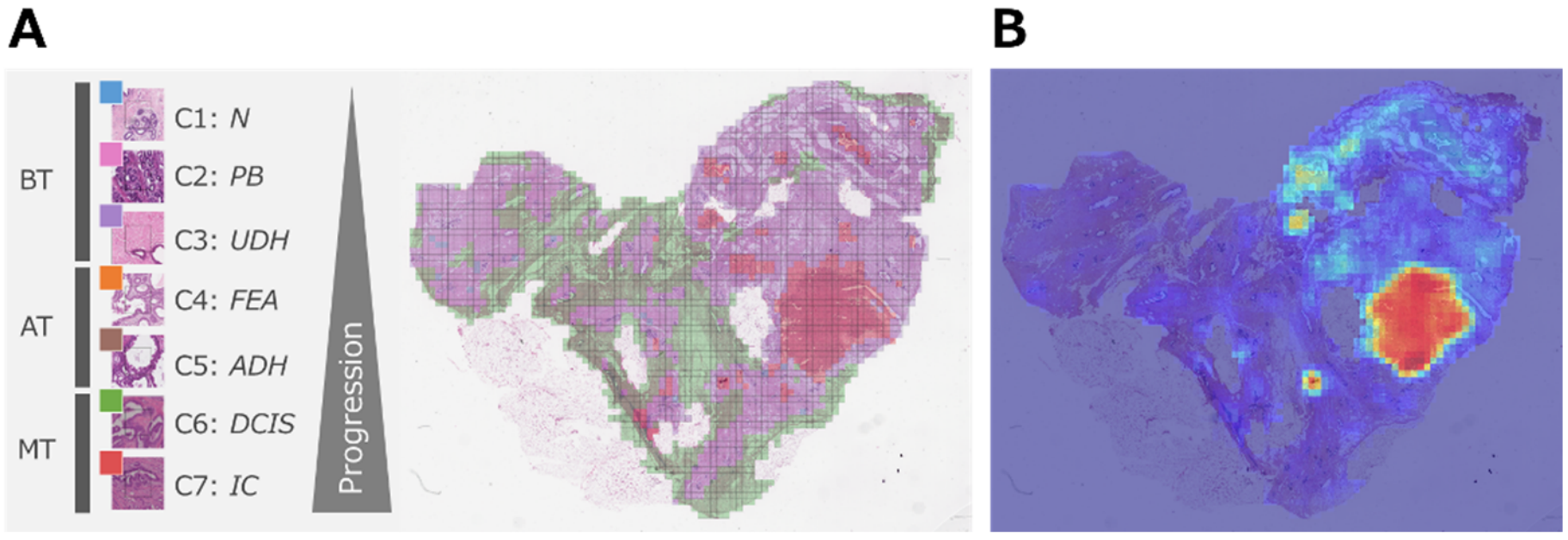

**Fig. 3.** Alignment between reference morphological prototypes and diagnostic relevance. (A) Patch-level mapping of supervised morphological concepts across the WSI. (B) Identification of regional features that drive the final diagnostic classification.

## Ablation Study

Table 2 shows the results of our ablation study. Compared to baselines utilizing solely spatial graph features ($\mathbf{H}_{graph}$) or prototype-aligned morphological features ($\mathbf{H}_{proto}$), our proposed residual fusion ($\mathbf{H}_{graph} + \mathbf{H}_{proto}$) substantially improves balanced accuracy while achieving comparable F1 performance to the prototype-only variant. These results support the benefit of combining structural context with morphological semantics is critical for robust WSI classification.

**Table 2.** Ablation study on the architecture components.

| Model Variant | Feature Usage | Bal. acc. | F1 |
|---|---|---|---|
| w/o Proto-Attn | $\mathbf{H}_{graph}$ (Graph features only) | 0.761±0.094 | 0.819±0.043 |
| w/o Residual | $\mathbf{H}_{proto}$ (Proto-aligned only) | 0.789±0.095 | 0.845±0.049 |
| Proposed | $\mathbf{H}_{graph}$ + $\mathbf{H}_{proto}$ (Residual Fusion) | **0.816±0.062** | **0.846±0.023** |

## Conclusion

We introduced ProtoGTX, a graph-based framework that combines spatial contextualization with a flexible morphological reference vocabulary for WSI analysis. Unlike approaches in which prototypes are optimized directly for the downstream prediction task, ProtoGTX separates the reference morphology from the predictive model and aligns graph-contextualized representations to this fixed vocabulary. Experiments on BRACS and CPTAC showed improved classification performance over the evaluated baselines, while qualitative analyses demonstrated more spatially coherent prototype maps and correspondence with pathologist annotations. This design enables morphological references to be derived from the training data or supplied externally, providing a flexible basis for semantic and spatial interpretation of WSI predictions. In this way, ProtoGTX provides a step toward bridging the gap between patch-level computation and slide-level pathological interpretation. Future work will investigate transferable reference vocabularies across pathological conditions and tissue domains.

## References

1. Lu, M. Y. *et al.* Data-efficient and weakly supervised computational pathology on whole-slide images. *Nat. Biomed. Eng.* **5**, 555–570 (2021).
2. Shao, Z. *et al.* TransMIL: Transformer based correlated multiple instance learning for whole slide image classification. *arXiv [cs.CV]* (2021).
3. Javed, S. A. *et al.* Additive MIL: Intrinsically interpretable multiple Instance Learning for pathology. *arXiv [cs.CV]* (2022).
4. Kapse, S. *et al.* SI-MIL: Taming Deep MIL for Self-Interpretability in Gigapixel Histopathology. *arXiv [cs.CV]* (2023).
5. Rymarczyk, D., Borowa, A., Tabor, J. & Zielinski, B. Kernel self-attention for weakly-supervised image classification using deep multiple instance learning. in *2021 IEEE Winter Conference on Applications of Computer Vision (WACV)* (IEEE, 2021). doi:10.1109/wacv48630.2021.00176.

6. Yang, S., Wang, Y. & Chen, H. MambaMIL: Enhancing long sequence modeling with sequence reordering in computational pathology. in *Lecture Notes in Computer Science* 296–306 (Springer Nature Switzerland, Cham, 2024).
7. Li, R., Yao, J., Zhu, X., Li, Y. & Huang, J. Graph CNN for survival analysis on whole slide pathological images. in *Medical Image Computing and Computer Assisted Intervention – MICCAI 2018* 174–182 (Springer International Publishing, Cham, 2018).
8. Ding, K. *et al.* Feature-enhanced graph networks for genetic mutational prediction using histopathological images in colon cancer. in *Lecture Notes in Computer Science* 294–304 (Springer International Publishing, Cham, 2020).
9. Zheng, Y. *et al.* A graph-transformer for whole slide image classification. *IEEE Trans. Med. Imaging* **41**, 3003–3015 (2022).
10. Hou, W. *et al.* H^2-MIL: Exploring hierarchical representation with heterogeneous multiple instance learning for whole slide image analysis. *Proc. Conf. AAAI Artif. Intell.* **36**, 933–941 (2022).
11. Chan, T. H., Cendra, F. J., Ma, L., Yin, G. & Yu, L. Histopathology whole slide image analysis with heterogeneous graph representation learning. in *2023 IEEE/CVF Conference on Computer Vision and Pattern Recognition (CVPR)* 15661–15670 (IEEE, 2023).
12. Li, J. *et al.* Dynamic graph representation with knowledge-aware attention for histopathology whole slide image analysis. *arXiv [cs.CV]* (2024).
13. Liu, J. *et al.* PAMIL: Prototype attention-based multiple instance learning for whole slide image classification. in *Lecture Notes in Computer Science* 362–372 (Springer Nature Switzerland, Cham, 2024).

14. Song, A. H. *et al.* Morphological prototyping for unsupervised slide representation learning in computational pathology. *arXiv [cs.CV]* (2024).

15. Sun, S., van Midden, D., Litjens, G. & Baumgartner, C. F. Prototype-based multiple instance learning for gigapixel whole slide image classification. in *Lecture Notes in Computer Science* 507–517 (Springer Nature Switzerland, Cham, 2026).

16. Rymarczyk, D. *et al.* ProtoMIL: Multiple Instance Learning with prototypical parts for whole-slide image classification. *arXiv [cs.LG]* (2021).

17. Zhao, B. *et al.* PTCMIL: multiple instance learning via prompt token clustering for whole slide image analysis. *Med. Image Anal.* **111**, 104013 (2026).

18. Yu, Z. *et al.* Structure-aware generalization for heterogeneous histopathology via prototype-based multiple instance learning. *NPJ Digit. Med.* **9**, 114 (2026).

19. Gao, Z., Zhang, Q., Guo, H., Wang, Y. & Li, Q. Non-parametric Prototypes Enable Semantic Graph Learning for Whole-Slide Pathology. in *Medical Image Computing and Computer Assisted Intervention -- MICCAI 2026* vol. 16891 342–352 (Springer Nature Switzerland, 2026).

20. Rigotti, M., Miksovic, C., Giurgiu, I., Gschwind, T. & Scotton, P. Attention-based interpretability with concept transformers. in *International conference on learning representations* (2021).

21. Edwards, N. J. *et al.* The CPTAC Data Portal: A resource for cancer proteomics research. *J. Proteome Res.* **14**, 2707–2713 (2015).

22. Brancati, N. *et al.* BRACS: A dataset for BReAst Carcinoma Subtyping in H&E histology images. *Database (Oxford)* **2022**, baac093 (2022).

23. Ma, J. *et al.* A generalizable pathology foundation model using a unified knowledge distillation pretraining framework. *Nat. Biomed. Eng.* https://doi.org/10.1038/s41551-025-01488-4 (2025) doi:10.1038/s41551-025-01488-4.

24. He, K., Zhang, X., Ren, S. & Sun, J. Deep residual learning for image recognition. *arXiv [cs.CV]* (2015) doi:10.48550/arXiv.1512.03385.

25. Chen, R. J. *et al.* Towards a general-purpose foundation model for computational pathology. *Nat. Med.* **30**, 850–862 (2024).